   \documentclass[final,3p,times,twocolumn]{elsarticle}

\usepackage{graphicx}
\usepackage[tight,footnotesize]{subfigure}
\usepackage[bookmarks]{hyperref}
\usepackage{amsmath,amssymb,amsfonts}

\journal{Computer Physics Communications}

\begin{document}

\begin{frontmatter}

\title{3D Magneto-Hydrodynamic Simulations of Parker Instability with Cosmic Rays}

\author[label1]{Ying-Yi Lo} 
  \ead{yylo@cycu.edu.tw}
\author[label2]{Chung-Ming Ko} 
  \ead{cmko@astro.ncu.edu.tw}
\author{Chih-Yueh Wang\corref{cor1}\fnref{label1}} 
 \ead{yueh@phys.cycu.edu.tw}
\address[label1]{Department of Physics,
Chung-Yuan Christian University,
Chungli, Taiwan 320, R.O.C.}
\address[label2]{Institute of Astronomy, 
Department of Physics and Center for Complex Systems, 
National Central University, Chung-Li,
Taiwan 320, R.O.C.
\\
{\rm Accepted by Computer Physics Communications(2010), doi:10.1016/j.cpc.2010.04.019,
Special Issue CCP 2009}
}

\begin{abstract}
This study investigates Parker instability in an interstellar medium (ISM) near the Galactic plane using three-dimensional magneto-hydrodynamic simulations. Parker instability arises from the presence of a magnetic field in a plasma, wherein the magnetic buoyant pressure expels the gas and cause the gas to move along the field lines. The process is thought to induce the formation of giant molecular clouds in the Galaxy. In this study, the effects of cosmic-ray (CR) diffusion are examined. The ISM at equilibrium is assumed to comprise a plasma fluid and a CR fluid at various temperatures, with a uniform magnetic field passing through it in the azimuthal direction of the Galactic disk. After a small perturbation, the unstable gas aggregates at the footpoint of the magnetic fields and forms dense blobs. The growth rate of the instability increases with the strength of the CR diffusion. The formation of dense clouds is enhanced by the effect of cosmic rays (CRs), whereas the shape of the clouds depends sensitively on the initial conditions of perturbation.
\end{abstract}


\end{frontmatter}



\section{Introduction}

 The gaseous disk of our Galaxy has been suggested to be unstable in response to perturbations in the magnetic field line that lies parallel to the Galactic plane. The dynamic process of the subsequent mixing of gas is called Parker instability in the astronomical literature (Parker 1966) [1]. The morphology of this instability can exhibit two modes - undular mode and interchange mode (Kim et al. 1998) [2]. The undular mode propagates along the magnetic field in the $x$ direction and is capable of exciting the gas in the vertical direction. An initially small perturbation in the undular mode can distort the uniform magnetic field and cause convection at full scales (Lee 2009) [3]. However, the interchange mode travels in the y direction, and may or may not affect the horizontal field lines. Hence, when the magnetic field has an oscillating component in the Galactic disk plane, convective instability may occur.

Once the gas rises because of magnetic buoyancy, the plasma can only fall along the magnetic field lines if the magnetic field is frozen into the plasma. Thus, a magnetic arch that protrudes from the bottom horizontal magnetic field is associated with dense footpoints. The key observation that supports Parker instability is that of giant molecular loops close to the Galactic center (Fukui et al. 2006) [4]. The gas aggregated at the footpoint is expected to collapse eventually, becoming a giant molecular cloud and a region that hosts stellar formations. The undular mode is also thought to power the magnetic dynamo, which sustain the magnetic field in the Galaxy (Hanasz et al. 2004) [5].

In this work, CR diffusion is adopted to elucidate Parker instability. Cosmic rays are highly energetic particles, 90\% of which are protons. The speed of individual CRs particles is close to that of light, but the bulk motion of CR is diffusive and the bulk speed of CRs is of the order of the Alfven speed. The CR particles propagate along the magnetic field lines as long as their gyroradius is significantly smaller than the characteristic spatial scales of the magnetic field. Cosmic rays may critically affect the dynamics of the ISM, since the energy density of CRs is of the same order as that of the magnetic field and turbulent gas motions. The importance of the effects of CRs has been acknowledged, and various works on CRs and Parker instability have been published. Hanasz et al (2004) identified the benefit of considering the CRs effect in the Galactic dynamo problem. Otmianowska-Mazur et al. (2009) [6] constructed a numerical model for the CR driven dynamo; they showed that excitement of fast growth in the magnetic flux and energy can account for the X-shaped magnetic structures of the galactic halos in several galaxies (Beck 2009) [7], which cannot be otherwise explained.

\section{Methodology}

A hydrodynamic approach is used to study the CR effect by constructing an MHD system that incorporates CRs as a second fluid. Cosmic rays are treated as a massless fluid, whose mass density is neglected, because the mass is negligible by comparison with the energy. The momentum spectrum of CR is neglected, simplifying the governing equations and making the computation much less intensive. The governing equations are, 

\begin{scriptsize}
\begin{eqnarray}
\frac{\partial \rho}{\partial t}&+& \bf{\nabla}\cdot ( \rho\bf{V})=\textit{0}, \label{parker
mass} \\
\frac{\partial \rho\textbf{V}}{\partial t}&+& \bf{\nabla}\cdot \left[ \rho \textbf{V} \textbf
{V}
+(\it{P_{g}}+\it{P_{c}}+\frac{B^2}{8\pi})\textbf{I}+\frac{\textbf{B}\textbf{B}}{4\pi}\right ]
 - \rho \bf{g}=0, \label{parker momentum} \\
\frac{\partial \textbf{B}}{\partial t}&+& \bf{\nabla}\times(\bf{V}\times\bf{B})=0, \label{par
ker induction}\\
\frac{\partial}{\partial t}(\frac{\gamma_{g}\it{P_{g}}}{\rho}&+&\frac{1}{2}\rho \it{v}^{2}
+\frac{\it{B}^2}{8\pi})\nonumber \\
&+&\bf{\nabla}{\cdot}\left[ (\frac{\gamma}{\gamma - 1} \it{P_{g}}
+\frac{1}{2}\rho \it{V}^{2})\bf{v}
+\frac{c}{4\pi}\bf{E}{\times}\bf{B}\right]\nonumber \\
&+&\bf{V}{\cdot} ({\nabla}\it{P_{c}}-\rho\bf{g})=0,\label{parker energe}\\
\frac{\partial}{\partial t}(\frac{P_c}{\gamma_{c}-1})&+& \bf{\nabla}\cdot (\frac{\gamma_c}
{\gamma_c -1}\it{P_{c}} ) \textbf{v}-\bf{v}\cdot\bf{\nabla}\it{P_c}\nonumber \\
&-&\bf{\nabla}\cdot\left[\overleftrightarrow{\kappa}\bf{\nabla}(\frac{\it{P_{c}}}{\gamma_{c}
-1})\right] =0.  \label{parker CR}
\end{eqnarray}
\end{scriptsize}
where 
$   \overleftrightarrow{\kappa} = (\kappa_{\|}-\kappa_{\bot}) \hat{b}\hat{b}
      -\kappa_{\bot}\delta_{ij} \label{kappa}
$
is the diffusion coefficient tensor,
$\rho$, $V$, $P_g$ and $\gamma_g$
are the mass density, velocity, pressure
and polytropic index of the plasma, respectively,
$B$ is the magnetic field, $g$ is the external gravity acceleration,
and $P_c$ and $\gamma_c$
are the pressure and adiabatic index of the CRs.
The CR pressure $P_c$ contributes energy to the plasma and modify the plasma momentum
according to its gradient and the magnetic field.
The magnetic field modifies the behavior of the plasma and
in return affect the CR via diffusion along the magnetic field lines.

The set of governing equations in Cartesian coordinates is solved. The modified Lax-Wendroff scheme is applied to the MHD part and the bi-conjugate gradients stabilized (BICGstab) method is applied to the diffusion part of the CR energy equation in the same manner as described elsewhere [8, 9]. The temperature distribution in a small rectangular region of the ISM is computed initially in hydrostatic equilibrium with the form 
\begin{equation}
C_{s}^{2}(z) = T(z)= T_{0} 
 +  (T_{halo}-T_{0})\frac{1}{2}[\tanh(\frac{z-z_{halo}}{w_{tr}})+1],
\label{temperature}
\end{equation}
where 
$T_0$ is the temperature of the Galactic disk; $T_{halo}$ and $z_{halo}$ are the temperature and 
height of the disk-halo interface, 
and $w_{tr}$ is the height of the transition between the disk and halo. 
The magnetic field is initially uniformly aligned in the $x$ direction, and independent of the height. Rotation of the disk is neglected. Given the initial temperature profile, the density, gas pressure, and magnetic pressure distribution at hydrostatic equilibrium are obtained by solving the ordinary differential equations. 
These values are then input to the simulations. A velocity perturbation with a one-dimensional sinusoidal wave-form is applied to the gas in a few grid zones on the horizontal plane. The most unstable wavelength identified in a previous linear analysis is employed to perturb the gas. In addition to wavelike perturbations, a point perturbation that represents a single supernova explosion is also applied. Supernova shock waves are thought to be efficient accelerators of particles with energies of up to $10^{15}$ eV. Calculations suggest that at least 10\% of the total supernova kinetic energy ($10^{51}$ ergs ) must be transformed into relativistic particles to generate the observed Galactic CRs. 10\% of the supernova energy is thus imparted to the CR energy. In the following,  the ratio of plasma pressure to magnetic pressure is denoted $\alpha$ and the ratio of plasma pressure to CR pressure is denoted $\beta$.

\section{Results and Conclusion}

 Parker instability causes gas to concentrate in a valley of magnetic field lines and collapse toward the equatorial plane. Previous investigations excluding CRs have demonstrated that if the external pressure is high, then plasma at the footpoints of the magnetic arches condenses to form either clumps or filaments, or becomes stabilized, depending on the velocity of rotation of the Galactic disk. The filaments tend to align in the $y$ direction, perpendicular to the magnetic field line; longer filaments are formed as the strength of the magnetic field increases, because the magnetic pressure supports the piling up of gas in the vertical direction. 

Various diffusion coefficients from $\kappa=10$ to 200 were considered. Figure 1(a) displays the case with $\alpha=1$, $\beta=10$ and $\kappa=2$. A comparison with the case without CRs (Fig.1(b)) shows that  the undular mode develops rapidly and the gas aggregates as spherical blobs. Linear analysis and 2D MHD simulations by Kuwabara et al. (2004) [9] indicated that the growth rate of Parker instability declines as the coupling between the CR and the gas becomes stronger (meaning that the CR diffusion coefficient becomes smaller). The MHD simulations herein verify this result. 

Cosmic rays like plasma gas also accumulate near the footpoint of the magnetic loop (Fig.2). 
For low $\kappa$ (stronger coupling between the two fluids), the CR pressure distribution is rather non-uniform, and a large CR pressure gradient force is exerted toward the top of the loop. This force impedes the falling motion of the gas to the bottom of the magnetic loop. Consequently, the growth rate of the instability is reduced. The growth rate normally increases with $\kappa$. However, the growth is maximal at $\kappa = 32$  - not 
$\kappa=200$. Furthermore, the change in the growth rate as $\kappa$ increases is small, because under strong diffusion ($\kappa > 40$), the speed at which matter falls along the magnetic field line may exceed the speed of sound as revealed by a shock close to the bottom of the magnetic loop, and so the CR is redistributed, causing the more uniform redistribution of the CR. Therefore, the contribution of CR to the gas dynamics is less important than in the low - CR pressure case. 

With respect to other parameters, increasing the transition height $w_{tr}$ dampens the undular mode but strengthens the interchange mode. However, if a short wavelength is applied to excite the instability, then the perturbation in the x-z plane induces the interchange mode, but without greatly affecting the undular mode. In the case of explosive perturbation, since the gas energy overwhelms the CR energy, instability develops rapidly at first, as if strong coupling existed between the CRs and gas. Nonetheless, the growth in later stages is similar to that which occurs using wavelike perturbations. 

The morphology of the unstable gas varies considerably depending on the form and the perturbation and the height at which it is applied. The 3D results herein are similar to those in 2D cases considered elsewhere [9] with a perturbation on the central x-z plane.

 The simulation results herein can be compared with the observed properties of nearby molecular clouds, such as the Taurus cloud, located at 140 pc from earth. This cloud is a good object for examining spontaneous cloud formation due to Parker instability, since the neighborhood contains no energy sources such as massive stars or supernova remnants that could trigger the collapse. Typical values of physical parameters in this region are evaluated. Speed of sound  $C = 5 \times  10^{4} \rm {cm/s}$
(corresponding to temperature T = 30 K), density 
$\rho =  2 \times 10^{21} \rm{g/cm^{3}}$
(corresponding to approximately 1000 particles per cubic centimeter), 
a scale height of 
$H = C /\sqrt{2\pi G \rho} = 1.7 \times 10^{18} \ \rm{cm} = 0.55 \ \rm{pc}$,
time units of 
$[t] = [H/C_s] = 3.5 \times 10^{13} \ \rm{sec} = 10^{6} \ \rm{yr}$,
and $\alpha=1$ are adopted, corresponding to a magnetic field strength of 10 $\mu G$. 
Observations 
have revealed that the Taurus cloud have two to three mutually parallel cylindrical regions, which is separated by a distance of 5 pc between. The distance between two filaments in 3D simulations that exclude CRs is approximately 10 H (Chou et al. 2000) [11], which value is consistent with a distance of ~5 pc. A similarly dense morphology has also been observed in the Ophiuchus cloud. Notably, the models herein do not show elongated condensation since only a one-dimensional sinusoidal wave was used to excite the instability. However, the slenderness of the filaments may also be attributed to a weak rotation or strong CR effect, which factors will be explored in future study.

{\bf Acknowledgement}
The authors would like to thank Jongsoo Kim and Wenchien Chou for useful correspondence and the National Science Council and the National Center for High-Performance Computing for supporting this work.


\begin{figure}[!t]
 \centering
     \subfigure[$\beta=10$]
   {
    \includegraphics[width=2.0in]{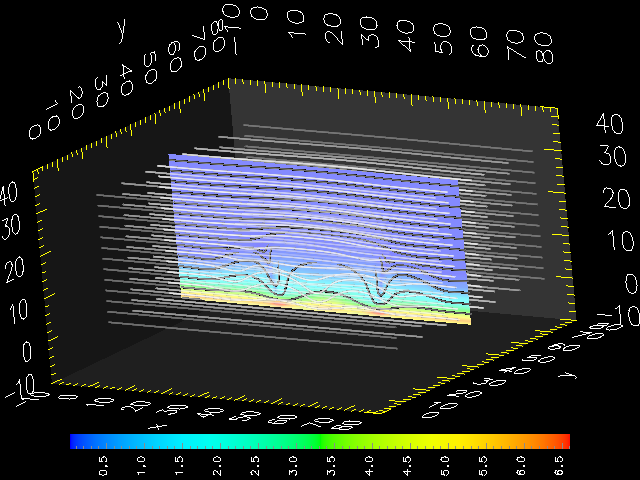}
  \label{mag:beta10}
  }
 \subfigure[$\beta \rightarrow \infty $]
  {
   \includegraphics[width=2.0in,angle=0]{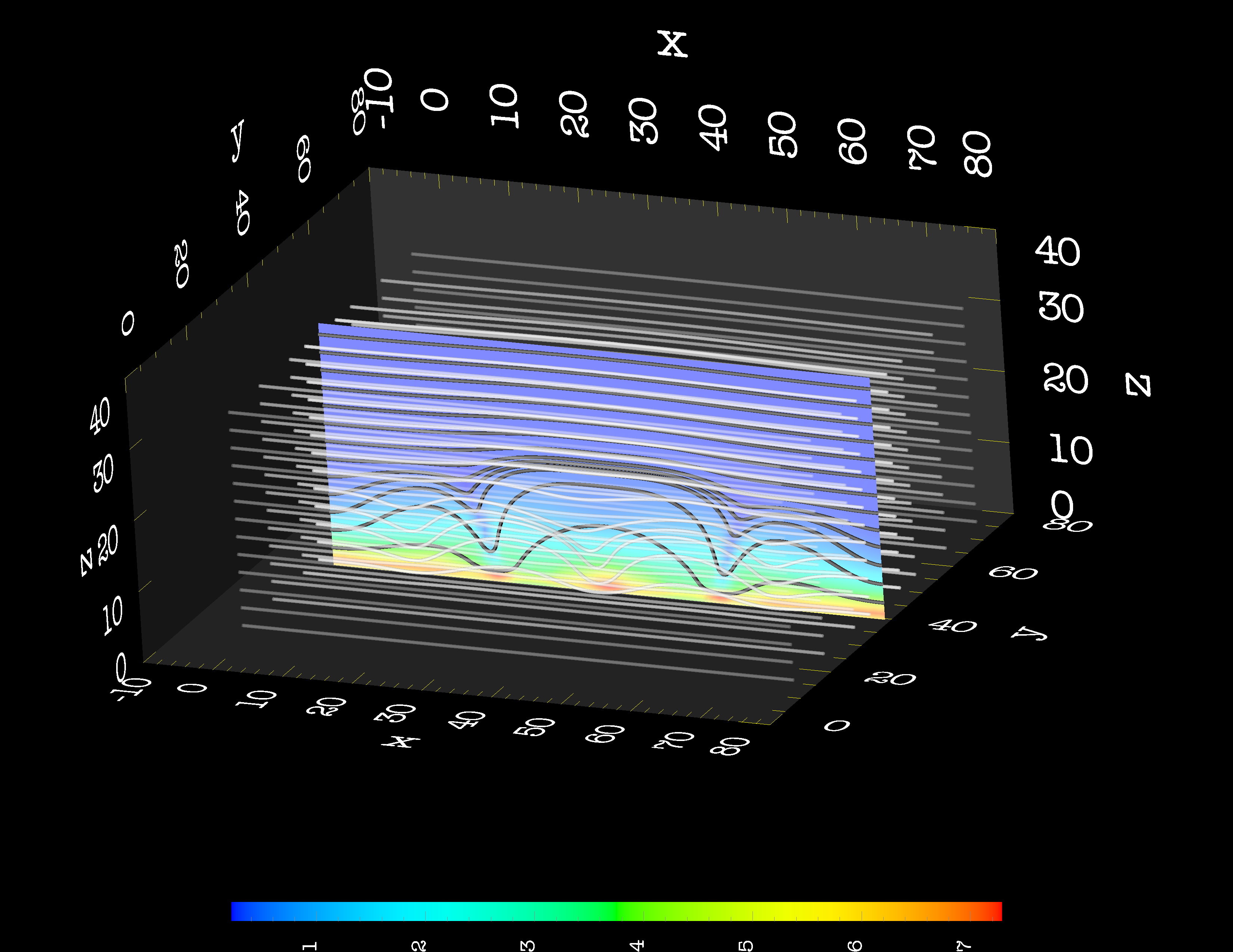}  
  \label{mag:cr00}
  }
 \caption{Contours of magnetic field lines for $\alpha=1$ and $\kappa_{\|}=2$}
 \end{figure}

%


\begin{figure}[!htbp]
\centering
\subfigure[$\log(\rho)$]
{
 \includegraphics[width=2.0in,angle=0]{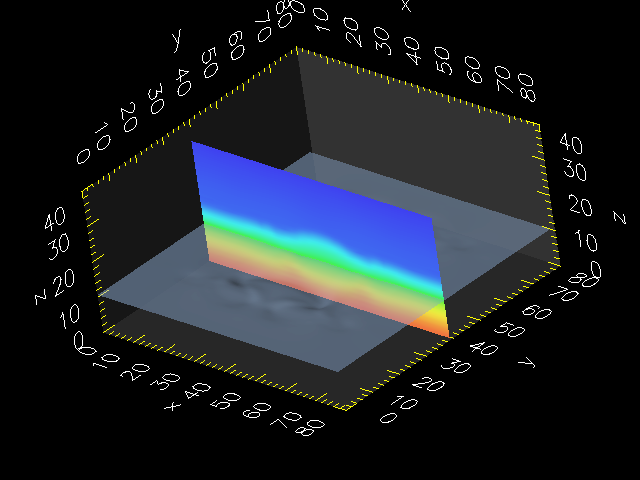}
\label{region2:absy < 3 over 2:rho}
}
\subfigure[$\log(P_g)$]
{
\includegraphics[width=2.0in,angle=0]{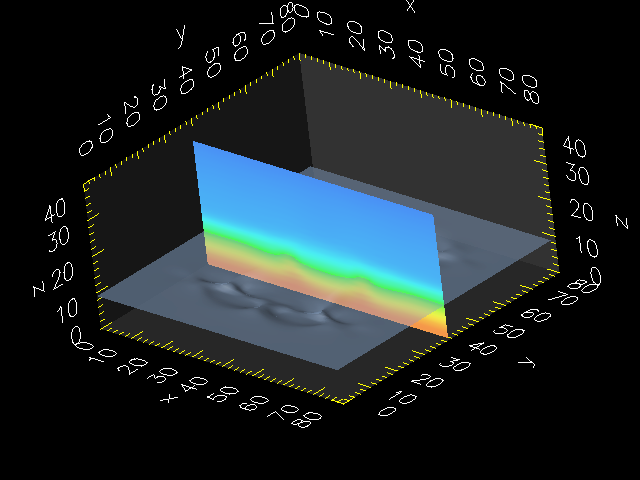}
\label{region2:absy < 3 over 2:pg}
}
\subfigure[$\log(P_c)$]
{
 \includegraphics[width=2.0in,angle=0]{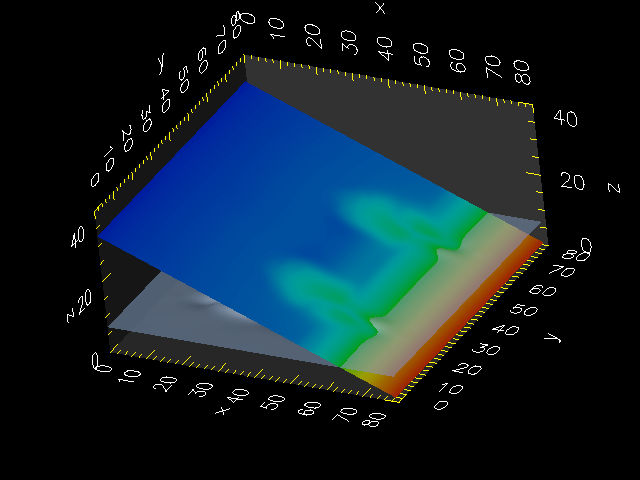}
\label{region2:absy < 3 over 2:pc}
}
\caption{Contours of mass density, gas pressure and cosmic-ray pressure}
\label{region2}
\end{figure}




\begin{thebibliography}{00}





\bibitem{parker66}
  E.~N.~Parker, Astrophysical Journal, 145 (1966), 881-833.

\bibitem{kim98}
  J.~Kim, S.~S.~Hong, D.~Ryu \& T.~W.~Jones, Astrophysical Journal, 506 (1998), L139-L142.

\bibitem{lee09}
  S.~E.~Lee, New Astronomy, 14 (2009) 44-50.

\bibitem{fukui06} 
 Y.~Fukui, et al., Science, 314 (2006), 106-109. 


\bibitem{hanasz04} 
 M.~Hanasz, G.~Kowal, K.~Otmianowska-Mazur and H.~Lesch, 
 Astrophysical Journal, 605 (2004), L33-L36.



\bibitem{otmianowska-mazur09}
 K.~Otmianowska-Mazur, M.~Soida, B.~Kulesza-\'Zydzik, M.~Hanasz \& G.~Kowal, Astrophysical Journal, 693 (2009), 1-7.


\bibitem{beck09}
R.~Beck, Astrophys. Space Sci. Trans., 5 (2009), 43-47.


\bibitem{yoko85}
 T. Yokoyama, K. Shibaba, PASJ 37 (1985) 31-46.

\bibitem{kuwabara04}
T.~Kuwabara, K. Nakamura \& C.~M.~Ko, Astrophysical Journal, 607 (2004), 828-839.

\bibitem{chou00}
 W.-S. Chou, R. Matsumoto, T. Tajima, M. Umekawa, K. Shibata, Astrophysical Journal 538 (2000) 710-727.


\end{thebibliography}



\end{document}